\documentclass[11pt]{article}
\usepackage{graphicx}
\usepackage[margin=1in]{geometry}
\usepackage{caption}
\usepackage{subcaption}
\usepackage{amsmath}
\usepackage{amssymb}
\usepackage{color}

\begin{document}

\begin{titlepage}
\begin{center}

{ \huge \bfseries The long-short wavelength mode coupling tightens primordial black hole constraints}\\[1cm]

Sam Young$^{1}$, Christian T. Byrnes$^{2}$\\[0.5cm]
$^{1,2}$Department of Physics and Astronomy, Pevensey II Building, University of Sussex, BN1 9RH, UK\\[0.5cm]
$^{1}$S.M.Young@sussex.ac.uk, $^{2}$C.Byrnes@sussex.ac.uk \\[1cm]

\today\\[1cm]

\end{center}

The effects of non-gaussianity on the constraints on the primordial curvature perturbation power spectrum from primordial black holes (PBHs) are considered. We extend previous analyses to include the effects of coupling between the modes of the horizon scale at the time the PBH forms and super-horizon modes. We consider terms of up to third order in the Gaussian perturbation. For the weakest constraints on the abundance of PBHs in the early universe (corresponding to a fractional energy density of PBHs  of $10^{-5}$ at the time of formation), in the case of gaussian perturbations, constraints  on the power spectrum are $\mathcal{P}_{\zeta}<0.05$ but can significantly tighter when even a small amount of non-gaussianity is considered, to $\mathcal{P}_{\zeta}<0.01$, and become approximately $\mathcal{P}_{\zeta}<0.003$ in more special cases. Surprisingly, even when there is negative skew (which naively would suggest fewer areas of high density, leading to weaker constraints), we find that the constraints on the power spectrum become tighter than the purely gaussian case - in strong contrast with previous results. We find that the constraints are highly sensitive to both the non-gaussianity parameters as well as the amplitude of super-horizon perturbations.

\end{titlepage}

\tableofcontents

\section{Introduction}
Theoretical arguments suggest that, if the right conditions are met, primordial black holes (PBHs) could have formed from the collapse of large density perturbations in the early universe. As a perturbation reenters the horizon, gravity can overcome the pressure forces and cause the perturbation to collapse to form a PBH with a mass of order the horizon mass. In order to collapse, then certain formation criteria need to be met, and this is normally stated in terms of the density contrast $\delta$ or the curvature perturbation $\zeta$. PBHs have traditionally been used to constrain the small scales of the early universe - and represent a unique window to constrain smallest scales. Whilst we have precision measurements from sources such as the cosmic microwave background (CMB) and large scale structure (LSS) (e.g.~the Planck constraints on inflation \cite{Ade:2013uln}), these only place constraints on a handful e-folds of the largest scales inside the visible universe. PBHs can be used to place constraints on the power spectrum spanning around 50 e-folds, although the constraints from PBHs are typically much weaker \cite{Josan:2009qn}. Ultra compact mini-haloes (UCMHs) can also be used to probe small scales \cite{Bringmann:2011ut}, although these constraints depend on dark matter particles decaying into observable particles, and do not cover as large a range of scales as the constraints from PBHs.

PBHs have never been observed, either directly or indirectly, but there are tight observational constraints on the abundance of PBHs, and these are used to constrain the power spectrum, as will be described later. The constraints on the abundance of PBHs come from the effects of their evaporation on the early universe for small PBHs, or the effects of their gravity on the later universe for larger ones. The constraints are typically stated in terms of $\beta$, the mass fraction of the universe going into PBHs at the time of formation. The constraints on $\beta$ range from $\beta\lesssim 10^{-25}$ to $\beta<10^{-5}$, depending on the mass of PBH being considered. For recent updates and a compilation of the constraints see \cite{Josan:2009qn, Carr:2009jm}.

The constraints on the power spectrum coming from PBHs are typically of order $10^{-2}$, orders of magnitude larger than those observed on cosmic scales. Whilst a spectral index less than unity, $n_{s}\approx0.96$, has been observed (e.g.~\cite{Ade:2013uln}) on cosmic scales, suggesting the power spectrum should become smaller on small scales, it is nonetheless possible for it to become large on small scales and form a significant number of PBHs. This can be seen in numerous models, including the running mass model \cite{Drees:2011hb}, axion inflation \cite{Bugaev:2013fya}, a waterfall transition during hybrid inflation \cite{Bugaev:2011wy, Lyth:2012yp,Halpern:2014mca}, from passive density fluctuations \cite{Lin:2012gs}, or in inflationary models with small field excursions but which are tuned to produce a large tensor-to-scalar ratio on large scales \cite{Hotchkiss:2011gz}. See also \cite{Linde:2012bt,Torres-Lomas:2014bua,Suyama:2014vga}. For further reading and a summary of various models which can produce PBHs, see \cite{Green:2014faa}. Alternatively, the constraint on the formation criteria can be relaxed during a phase transition in the early universe, causing PBHs to form preferentially at that mass scale, e.g.~\cite{Jedamzik:1999am}.

The constraints from PBHs on the primordial power spectrum are highly sensitive to even small amounts of non-gaussianity, and this has been studied extensively in the literature (e.g.~\cite{Bullock:1996at,Ivanov:1997ia,Seery:2006wk,Shandera:2012ke}), and in this paper we extend the calculation conducted by Byrnes, Copeland, Green and Young \cite{Byrnes:2012yx, Young:2013oia} to include the effects of large scale inhomogeneities in the distribution caused by non-gaussianity. 

In Section 2, we review how constraints on the abundance of PBHs can be used to constrain the power spectrum, and in Section 3 we review previous calculations how local-type non-gaussianity affects these constraints, as well as a more general discussion of the effects of non-gaussianity. In Section 4, we describe how the presence of non-gaussianity and large super-horizon modes can affect the abundance of PBHs which form on smaller scales, and apply this to the calculation of constraints in Sections 5 and 6, for quadratic and cubic type non-gaussianity respectively. We finish with a discussion of key points in Section 7.

\section{Constraining the power spectrum}

Using the fact that PBHs have not been observed, one can place an upper limit on the primordial power spectrum on scales which could not otherwise be constrained. In this paper, this upper limit on the power spectrum, and its dependance upon non-gaussianity, will be calculated. There are different constraints on the abundance of PBHs of different masses - and therefore different constraints on the primordial power spectrum \cite{Josan:2009qn}.

The abundance of PBHs is normally stated as the mass fraction of the universe contained within PBHs at the time of formation, $\beta$, and in a recent paper we showed how this can be calculated directly from the curvature perturbation power spectrum, $\mathcal{P}(\zeta)$, matching well with the traditional calculation (which calculates the abundance by using window functions to smooth the distribution). $\beta$ is given by
\begin{equation}
\beta=2\int_{\zeta_{c}}^{\infty}P(\zeta)d\zeta,
\end{equation}
where $\zeta_{c}$ is the threshold value for PBH formation, and $P(\zeta)$ is the probability density function (PDF) of $\zeta$. In the case of a gaussian distribution, this can be approximated as \cite{Byrnes:2012yx}
\begin{equation}
\beta=\mathrm{erfc}\left(\frac{\zeta_{c}}{\sqrt{2}\sigma}\right)\approx\mathrm{exp}\left(-\frac{\zeta_{c}^{2}}{2\sigma^{2}}\right).
\end{equation}
This can be rewritten to show how the constraints on $\beta$ give constraints on $\mathcal{P}_{\zeta}$,
\begin{equation}
\mathcal{P}_{\zeta}=\sigma^{2}=\sqrt{\frac{\zeta_{c}^{2}}{2\ln\left(1/\beta\right)}}.
\end{equation}
In this paper, we will take the threshold value for PBH formation to be $\zeta_{c}=1$ \cite{Shibata:1999zs,Green:2004wb}\footnote{In order to be consistent with calculations using the density contrast, it is preferable to use a larger value, $\zeta_{c}\approx1.2$ (the upper value found in \cite{Shibata:1999zs}), which matches better with the expected critical value of the density contrast, $\Delta_{c}\approx0.5$. However, whilst $\beta$ is extremely dependant on $\zeta_{c}$, the constraints on the power spectrum do not change significantly - and we use $\zeta_{c}=1$ in order to be consistent with previous papers.}. Significant uncertainty on  the critical value of collapse remains and the result depends on the density profile \cite{Musco:2004ak,Hidalgo:2008mv,Musco:2008hv,Nakama:2013ica,Harada:2013epa,Nakama:2014fra}. For $\beta<10^{-5}$ and $\beta<10^{-20}$, for a gaussian distribution this gives the constraints $\mathcal{P}_{\zeta}<0.0513$ and $\mathcal{P}_{\zeta}<0.0115$ respectively.

\begin{figure}[t]
\centering
\begin{subfigure}{\textwidth}
\includegraphics[width=\linewidth]{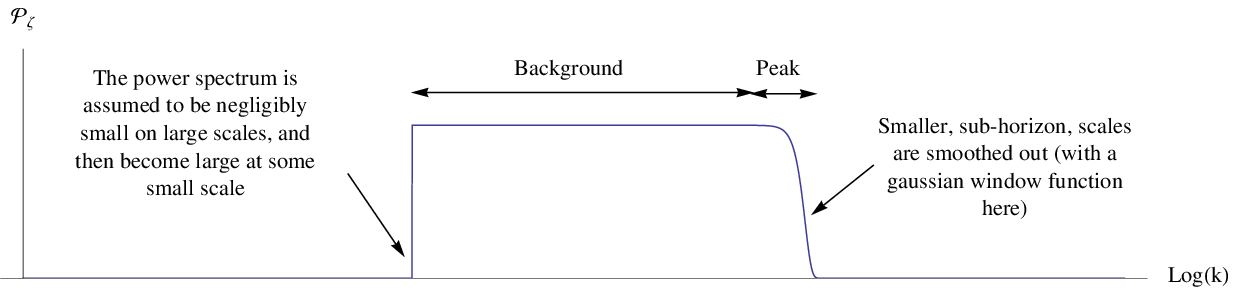}
\end{subfigure}
\begin{subfigure}{\textwidth}
\includegraphics[width=\linewidth]{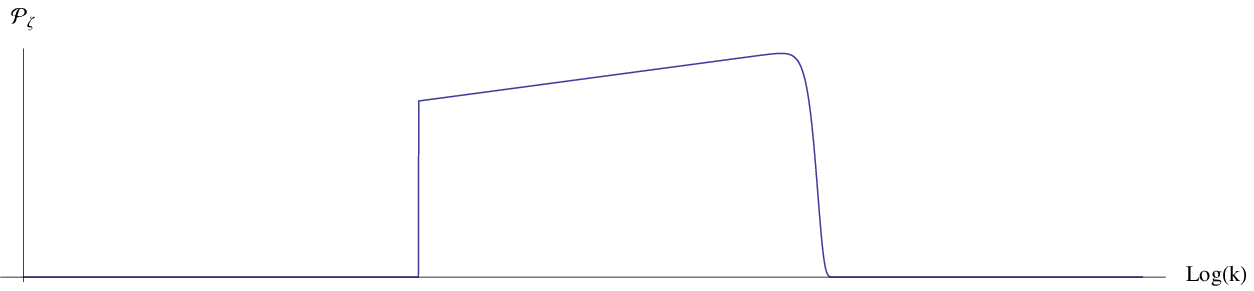}
\end{subfigure}
\caption{The form of the power spectrum being used in this paper is shown. For simplicity, we assume that on large scales the power spectrum is negligibly small, $\mathcal{P}_{\zeta}\ll 1$, before quickly becoming large at some scale (in this case with a step function). The power spectrum is then assumed to be large down to arbitrarily small scales - although the effect of smoothing reduces the power on sub-horizon scales to be effectively zero. The 'peak' scales correspond to the scale at which PBHs are forming at a given time (the horizon scale), where the 'background' scales are so large as to be unobservable. The top figure shows a flat spectrum, which is assumed to be the case for a gaussian distribution. However, for a non-gaussian distribution, the effect of coupling between modes will typically serve to increase the power on small scales, even when the amplitude of the Gaussian perturbations is scale invariant, as shown in the bottom figure.}
\label{powerSpectrum}
\end{figure}

In previous papers, we used PBH constraints to calculate how the constraints on the power spectrum depend on the amount of non-gaussianity present (see section 3), in the local model of non-gaussianity \cite{Byrnes:2012yx, Young:2013oia}. In this paper, we go beyond previous calculations, and account for large scale inhomogeneities in the power spectrum caused by the non-gaussian terms as documented in \cite{Byrnes:2011ri}. Whilst large super-horizon modes in the curvature perturbation do not affect the local evolution of the universe and therefore do not affect whether a region collapses to form a PBH or not \cite{Young:2014ana}, they can have an indirect effect due to their influence on smaller scale modes. In this paper, we will assume that the power spectrum becomes large below a certain scale (as demonstrated in Fig.~\ref{powerSpectrum}), and place constraints on the amplitude of this power spectrum from the constraints on the abundance of PBHs. The top power spectrum shown in Fig.~\ref{powerSpectrum} is scale invariant - which we assume to be the case for a gaussian distribution. However, for a non-gaussian distribution, the power spectrum increases as $k$ increases, which is due to the effects of modal coupling - so even though the gaussian component of the perturbations is constant, overall the power spectrum increases. For a specific model, such a power spectrum is unlikely and a more suitable model for the power spectrum should be used.

\section{Review of non-gaussian constraints}

It has previously been shown that the constraints which can be placed on the power spectrum depend upon the distribution present in the early universe (recent papers include \cite{Shandera:2012ke,Byrnes:2012yx,Young:2013oia,Bugaev:2012ai}), and that the mass fraction of the early universe going into PBHs, $\beta$, is strongly dependant on the amount of non-gaussianity present. 

\begin{figure}[t]
\centering
	\begin{subfigure}{0.5\textwidth}
	\centering
	\includegraphics[width=\linewidth]{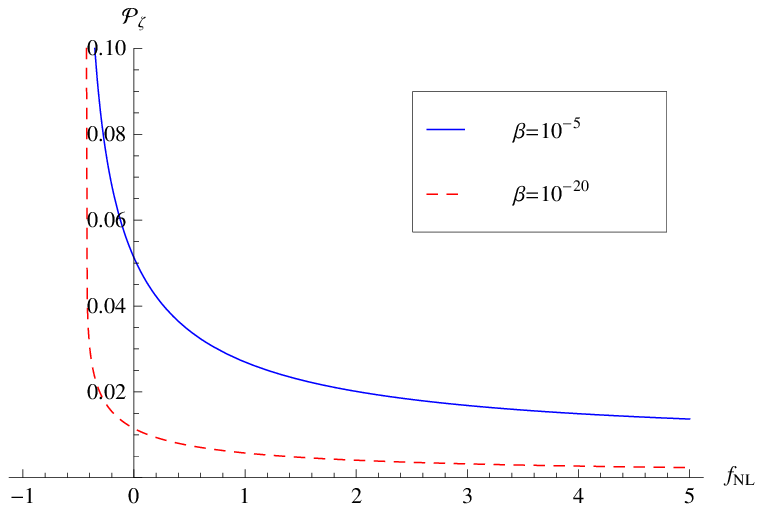}
\end{subfigure}%
\begin{subfigure}{0.5\textwidth}
	\centering
	\includegraphics[width=\linewidth]{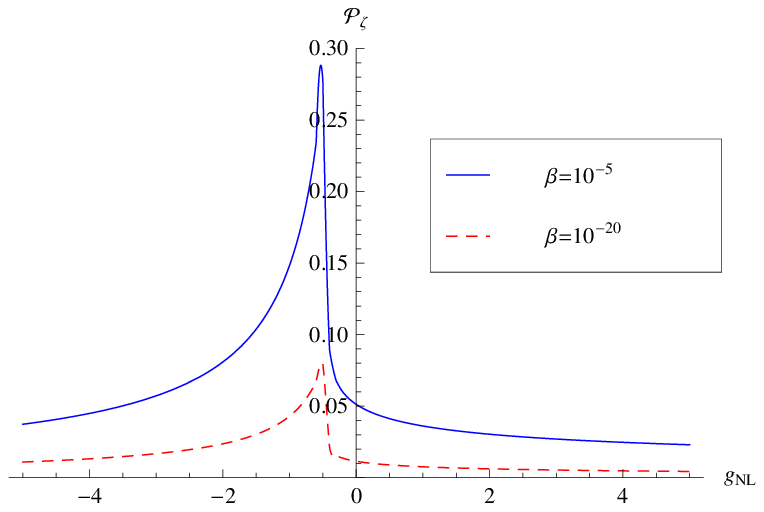}
\end{subfigure}
 \caption{In the local model of non-gaussianity, the constraints on the power spectrum, $\mathcal{P}_{\zeta}$, depend strongly upon the non-gaussianity parameters. The left plot shows how constraints depend on $f_{NL}$ (assuming all higher order terms are zero). The constraints tighten significantly for positive $f_{NL}$ but weaken dramatically for negative $f_{NL}$. The right plot shows how constraints depend on $g_{NL}$ (assuming all higher order terms and the quadratic term are zero). For most values of $g_{NL}$ the constraints are tighter than the gaussian case, but significantly weaker for small negative values of $g_{NL}$.}
\label{old constraints}
\end{figure}

In this paper, we will consider the local model of non-gaussianity to third-order,
\begin{equation}
\zeta=\zeta_{G}+\frac{3}{5}f_{NL}\left(\zeta_{G}^{2}-\sigma^{2}\right)+\frac{9}{25}g_{NL}\zeta_{G}^{3}=h\left(\zeta_{G}\right),
\end{equation}
where $\sigma^{2}=\langle\zeta_{G}^{2}\rangle$. We define the solution to this equation as $\zeta_{G}=h^{-1}(\zeta)$, and $\beta$ can be expressed in terms of $h^{-1}(\zeta)$ \cite{Byrnes:2012yx}. Note that, whilst the meaning of $f_{NL}$ and $g_{NL}$ in this paper are the same as that used in observational cosmology of CMB and LSS, similar values of these parameters here have a much larger effect on the distribution than in the CMB or LSS. This is because the constraint on the amplitude of perturbations is much weaker - typically of order $10^{-1}$ rather than $10^{-5}$. Therefore, $f_{NL}\approx 1$ represents approximately a $10\%$ correction. We will here briefly review previous work by considering the case of positive $f_{NL}$ and zero $g_{NL}$, $h^{-1}(\zeta)$ has two solutions, given by
\begin{equation}
h_{\pm}^{-1}(\zeta)=\frac{-5\pm\sqrt{25+36f_{NL}^{2}\sigma^{2}+60f_{NL}\zeta}}{6f_{NL}}.
\end{equation}
$\beta$ can then be calculated by integrating over the PBH forming values of $\zeta_{G}$, giving \footnote{This is equivalent to integrating over the probability distribution function of $\zeta$: $\beta=2\int_{\zeta_{C}}^{\infty}P(\zeta)d\zeta$.}
\begin{equation}
\beta=\mathrm{erfc}(h_{+}^{-1}(\zeta_{c}))+\mathrm{erfc}(\lvert h_{-}^{-1}(\zeta_{c})\rvert).
\label{quadraticbeta}
\end{equation}
The full derivation can be seen in \cite{Byrnes:2012yx}. This expression can then be solved numerically for a given constraint on $\beta$, such as $\beta<10^{-5}$, to find a constraint on $\sigma$, and a constraint on the power spectrum can be calculated using \cite{Byrnes:2007tm}
\begin{equation}
\mathcal{P}_{\zeta}=\sigma^{2}+4\left(\frac{4f_{NL}}{5}\right)^{2}\sigma^{4}\mathrm{ln}(kL).
\end{equation}
Fig.~\ref{old constraints} shows how the constraints on the power spectrum depend upon the non-gaussianity parameters $f_{NL}$ and $g_{NL}$ for $\beta=10^{-5}$ and $\beta=10^{-20}$. 
\begin{itemize}
\item{The $f_{NL}$ term affects the skew of the distribution - a positive $f_{NL}$ enhances the tail of the distribution, increasing PBH production, which means the constraints become tighter. For negative $f_{NL}$, the constraints weaken dramatically. There is a maximum value of $\zeta$ given by
\begin{equation}
\zeta_{max}=-\frac{5}{6f_{NL}}+\frac{3}{5}f_{NL}\left(\frac{25}{36 f_{NL}^{2}}-\sigma^{2}\right),
\end{equation}
which is a function of $\sigma$. In order for any PBHs to form, $\zeta_{max}$ must be greater than $\zeta_{c}$, and so for $f_{NL}<-\frac{5}{12}$, $\sigma$ must be above a certain value $\sigma_{c}$,
\begin{equation}
\sigma_{c}=\frac{\sqrt{-25-60f_{NL}}}{6f_{NL}}.
\end{equation}
If $\sigma$ (and so the power spectrum) is below this value, no PBHs are formed, but typically, if $\sigma$ is larger then too many PBHs form. This means that an extreme fine tuning of the power spectrum is required in order to generate a small but non-zero amount of PBHs.}
\item{The $g_{NL}$ term affects the kurtosis of the distribution. For positive $g_{NL}$. the tails of the probability density function are enhanced - meaning tighter constraints. For small negative values, the tails are diminished - meaning weaker constraints - but as $g_{NL}$ becomes more negative the tails become more enhanced - meaning constraints again become tighter.}
\end{itemize}

Similar behaviour is displayed for higher order terms - even terms have a similar effect as the quadratic term, and odd order terms have a similar effect to the cubic term. The effects of combining higher order terms was investigated \cite{Young:2013oia}, finding that for certain models displaying a simple relation between the non-gaussianity parameters ($g_{NL}\propto f_{NL}^{2}$, $h_{NL}\propto f_{NL}^{3}$) the constraints calculated converge, but that care should be taken as this might not always be the case. 

\section{Large-scale inhomogeneities from non-gaussianity}

In this section, we describe how the presence of local non-gaussianity leads to a coupling between long and short wavelength modes, and thus how a mode which is greatly super-horizon at the time of PBH formation can have an effect on the distribution of PBHs on smaller scales. For a more detailed calculation and discussion of implications, the reader is directed to \cite{Byrnes:2011ri}.

\begin{figure}[t]
\centering
\begin{subfigure}{0.8\textwidth}
	\centering
	\includegraphics[width=\linewidth]{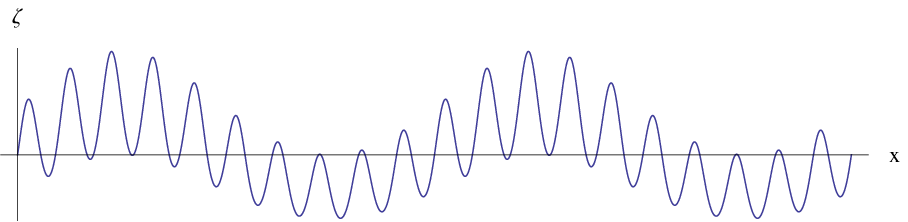}
\end{subfigure}
\begin{subfigure}{0.8\textwidth}
	\centering
	\includegraphics[width=\linewidth]{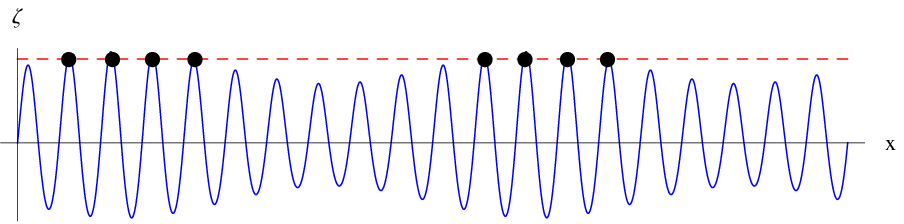}
\end{subfigure}
 \caption{The first (top) figure shows one arbitrary realisation of a universe containing exactly one long wavelength and one short wavelength gaussian mode, and the corresponding non-gaussian components where the universe contains quadratic non-gaussianity. At the time when the short wavelength mode reenters the horizon after the end of inflation, the long wavelength mode is not yet visible - and will not affect the local evolution of the universe (i.e.~whether it forms a PBH or not). The second (bottom) plot shows the same universe with the long wavelength mode subtracted. $\zeta$ can now be used as a formation criterion for the formation of PBHs - if it is over a certain value, then that region will collapse to form a PBH. The dashed red line shows such a formation criterion, and the black circles represent areas which will collapse to form a PBH.}
\label{large scale inhomos}
\end{figure}

We will consider a universe with a distribution in $\zeta$ described by the local model of non-gaussianity (equation (\ref{quadratic local model})), but which contains exactly 2 gaussian modes. We can therefore decompose the gaussian component of $\zeta$ into its two components
\begin{equation}
\zeta_{G}=\zeta_{s}+\zeta_{l}.
\end{equation}
The first plot in Fig.~\ref{large scale inhomos} shows one possible realisation of such a universe, with 2 gaussian modes of arbitrary size. In this picture, the non-gaussian components to not appear to be very important - they are small corrections to the existing gaussian components.  However, as described in \cite{Young:2014ana}, super-horizon modes should not be considered when deciding if a region will collapse to form a PBH. We will study the time at which PBHs form on the scale of the shorter-scale mode (when that mode enters the horizon), and therefore neglect the components of $\zeta$ which depend only on the long wavelength mode. The second plot in figure\ref{large scale inhomos} shows the relevant modes for formation of PBHs: the red dashed line represents a hypothetical formation criterion for PBHs and the black dots represent PBH forming regions. We note that in certain regions of the universe corresponding to peaks in the super-horizon mode, PBHs are produced in significant numbers, whilst in regions corresponding to troughs in the super-horizon mode, no PBHs would be produced.

The effect of different scale modes on the formation of primordial black holes has recently been investigated by Nakama \cite{Nakama:2014fra}. Nakama investigated the case where a large perturbation which will collapse to form a PBH is itself superposed on a much larger perturbation which will also collapse to form a PBH upon reentry. The smaller PBH, which forms first, is swallowed by the second PBH as it forms, leading to a single large PBH. As expected, the first collapse is unaffected by the large scale perturbation as it is outside the horizon at the time of collapse, and the second collapse is unaffected by the first due to the large scale difference between the two. Nakama also investigates the effect of sub-horizon modes on the possible collapse of a perturbation, finding that the presence of such modes lowers the threshold value for collapse - making the collapse of such a perturbation more likely. This a separate effect to the one which we are investigating in this paper - here, the effect of super-horizon modes on the distribution of horizon-scale perturbations is studied, whilst Nakama describes the effect of sub-horizon modes on the evolution of horizon-scale perturbations. The net result of the sub-horizon modes is to lower the formation threshold for PBHs, which would serve to further tighten the constraints derived in this paper.

\section{Inhomogeneous quadratic non-gaussianity}
In the local model of non-gaussianity, the curvature perturbation $\zeta$ is given to $2^{nd}$ order by
\begin{equation}
\label{quadratic local model}
\zeta=\zeta_{G}+\frac{3}{5}f_{NL}\left(\zeta_{G}^{2}-\langle\zeta_{G}^{2}\rangle\right),
\end{equation}
where $\zeta_{G}$ is a gaussian variable, and it is necessary to subtract the $\langle\zeta_{G}^{2}\rangle$ term in the above expression so that the expectation value of $\zeta$ remains zero, $\langle\zeta\rangle=0$.

We will now use the peak-background split, separating the gaussian component of the curvature perturbation $\zeta_{G}$ into a large scale  'background' perturbation $\zeta_{l}$ and a small scale 'peak' perturbation $\zeta_{s}$,
\begin{equation}
\zeta_{G}=\zeta_{l}+\zeta_{s}.
\end{equation}
The full expression for the curvature perturbation $\zeta$ then becomes
\begin{equation}
\zeta=\left(\zeta_{l}+\zeta_{s}\right)+\frac{3}{5}f_{NL}\left(\left(\zeta_{l}+\zeta_{s}\right)^{2}-\langle\left(\zeta_{l}+\zeta_{s}\right)^{2}\rangle\right).
\end{equation}
Terms which are independent of $\zeta_{s}$, and depend only on the large scale perturbation $\zeta_{l}$ can be neglected - as they are not visible at the time of PBH formation, leaving
\begin{equation}
\label{quadratic inhomogeneity}
\zeta=\left(1+\frac{6}{5}f_{NL}\zeta_{l}\right)\zeta_{s}+\frac{3}{5}\left(\zeta_{s}^{2}-\sigma_{s}^{2}\right).
\end{equation}
In a small patch of the universe, $\zeta_{l}$ will appear constant, and the above expression can be written in terms of new variables $\tilde{\zeta}_{G}$, $\tilde{\sigma}$ and $\tilde{f}_{NL}$, given by
\begin{equation}
\tilde{\zeta}_{G}=\left(1+\frac{6}{5}f_{NL}\zeta_{l}\right)\zeta_{s},
\end{equation}
\begin{equation}
\tilde{\sigma}=\left(1+\frac{6}{5}f_{NL}\zeta_{l}\right)\sigma_{s},
\end{equation}
\begin{equation}
\tilde{f}_{NL}=\left(1+\frac{6}{5}f_{NL}\zeta_{l}\right)^{-2}f_{NL}.
\end{equation}
This allows equation (\ref{quadratic inhomogeneity}) to be written in a form analogous to equation (\ref{quadratic local model}),
\begin{equation}
\zeta=\tilde{\zeta}_{G}+\frac{3}{5}\tilde{f}_{NL}\left(\tilde{\zeta}_{G}^{2}-\tilde{\sigma}^{2}\right)=\tilde{h}(\tilde{\zeta}_{G}).
\label{localQuadInhomo}
\end{equation}
Taking $\zeta_{l}$ to be constant in a given region of the universe, the mass fraction of the region going into PBHs $\tilde{\beta}$ can then be written in terms of the locally observable values $\tilde{f}_{NL}$, $\tilde{\zeta}_{G}$ and $\tilde{\sigma}$ in the same way as in equation (\ref{quadraticbeta}):
\begin{equation}
\tilde{\beta}=\mathrm{erfc}(\tilde{h}_{+}^{-1}(\zeta_{c}))+\mathrm{erfc}(\lvert \tilde{h}_{+}^{-1}(\zeta_{c})\rvert).
\end{equation}
However, this is still a function of $\zeta_{l}$, and to obtain the mass fraction of the entire universe going into PBHs, this should be integrated over $\zeta_{l}$
\begin{equation}
\label{quadratic beta}
\beta=\int_{-\infty}^{\infty}\tilde{\beta}(\zeta_{l})P(\zeta_{l})d\zeta_{l},
\end{equation}
where $P(\zeta_{l})$ is the (gaussian) PDF of $\zeta_{l}$. Therefore, $\beta$ depends not only on the variance (power spectrum) of the small scale perturbations (which is the scale PBH formation occurs at), but also on the variance of the large scale modes. In this paper, we assume the form of the power spectrum shown in Fig.~\ref{powerSpectrum}  - and therefore, the variance of the large scale perturbations can be written as a function of the variance of the small scale perturbations, depending on the number of e-folds one considers.

The variance of the large scale perturbations is given by integrating the power spectrum multiplied by a smoothing function $W(k R)$, where $R$ is the smoothing scale, as follows
\begin{equation}
\langle\zeta_{l}^{2}\rangle=\int_{0}^{\infty}d\ln(k)W^{2}(k R)\mathcal{P}_{\zeta_{l}}(k).
\end{equation}
In practice, since we are assuming a scale invariant power spectrum (for the gaussian components), which is zero below a certain value of $k$, then $\langle\zeta_{l}^{2}\rangle$ depends upon the number of e-folds $\mathcal{N}$ considered to be part of the 'background' large scale perturbation. We will approximate that
\begin{equation}
\label{efolds and variance}
\sigma_{l}=\sqrt{\langle\zeta_{l}^{2}\rangle}\approx \sqrt{\mathcal{N}} \sigma_{s},
\end{equation}
in order to derive constraints on the power spectrum from the constraints on the abundance of PBHs. Equation (\ref{efolds and variance}) can be substituted into equation (\ref{quadratic beta}), which can then be solved numerically to find a constraint on $\sigma_{s}$ from a constraint on $\beta$. The constraint on the power spectrum $\mathcal{P}_{\zeta}$ can then be calculated using \cite{Boubekeur:2005fj,Byrnes:2007tm}
\begin{equation}
\mathcal{P}_{\zeta}=\sigma_{s}^{2}+4\left(\frac{3}{5}f_{NL}\right)^{2}\sigma_{s}^{4}\ln\left(kL\right),
\end{equation}
where the cut-off scale $L\approx\frac{1}{H}$ is of order the horizon-scale, $k$ is the scale of interest. The factor $\ln\left(kL\right)$ can therefore become significant, as the power spectrum is taken to be large across a number of e-folds - and will be approximately equal to the number of e-folds being considered, $\mathcal{N}$ \cite{Suyama:2008nt,Kumar:2009ge}.

\begin{figure}[t]
\centering
\begin{subfigure}{0.5\textwidth}
	\centering
	\includegraphics[width=\linewidth]{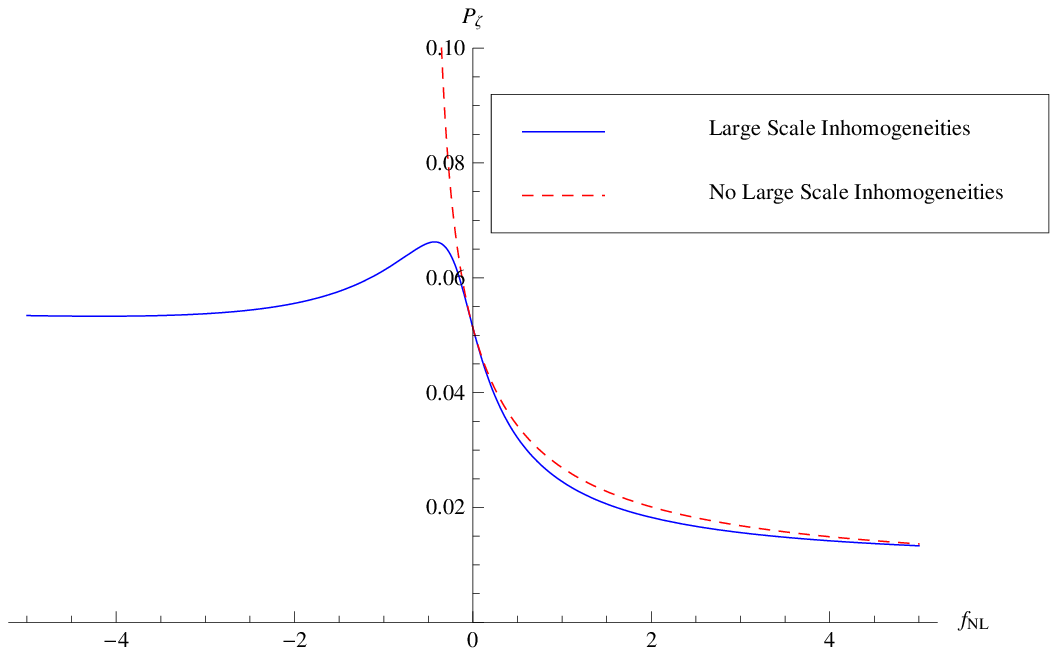}
	\caption{$\beta=10^{-5}$}
\end{subfigure}%
\begin{subfigure}{0.5\textwidth}
	\centering
	\includegraphics[width=\linewidth]{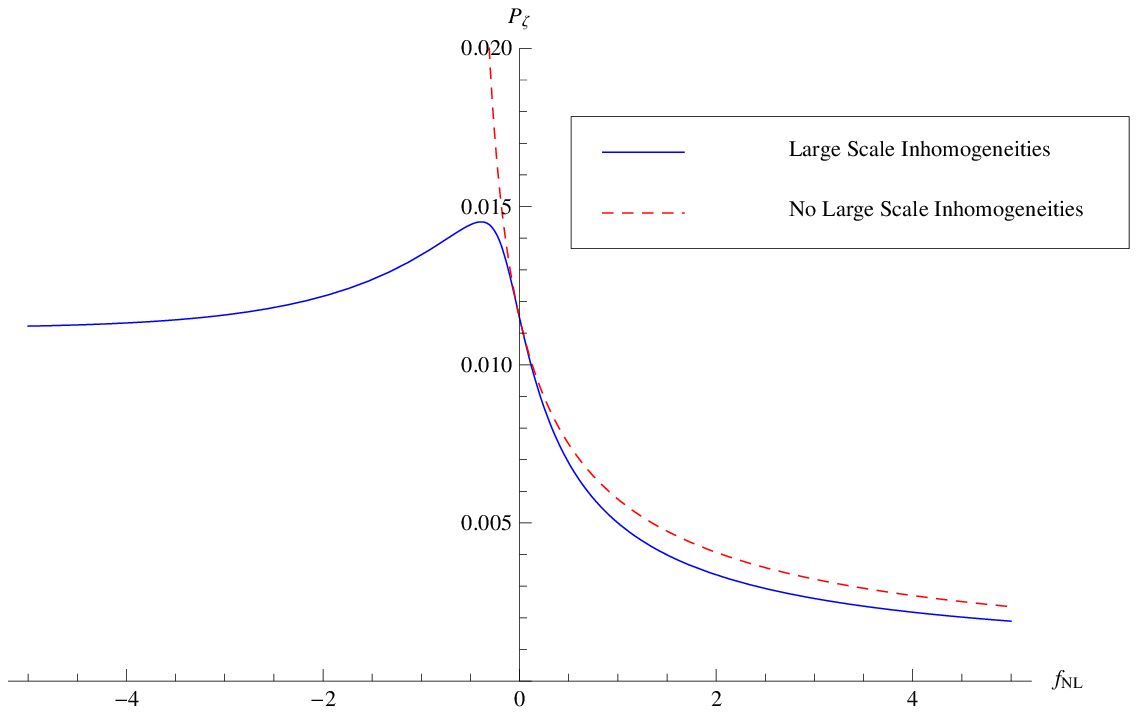}
	\caption{$\beta=10^{-20}$}
\end{subfigure}
 \caption{The constraints that can be placed upon the power spectrum are displayed - they depend significantly upon the value of the non-gaussianity parameter, $f_{NL}$. The dotted red lines show the constraints calculated previously, where the effect of large scale modes was not considered, and the solid blue lines show the constraints when they are included. The left plot (a) displays the constraints for $\beta<10^{-5}$ and the right plot (b) for $\beta<10^{-20}$. In this plot it is assumed that the variance of the gaussian component of the large scale perturbations is the same as that of the small scale perturbations, $\langle\zeta_{l}^{2}\rangle=\langle\zeta_{s}^{2}\rangle=\sigma^{2}$. For positive $f_{NL}$ the constraints are tighter than the gaussian case, and slightly stronger than in previous calculations ignoring modal coupling. For negative $f_{NL}$, the constraints are similar to the gaussian case, and the dramatic weakening of the constraints as $f_{NL}$ becomes negative is no longer seen.}
\label{fnl constraints}
\end{figure}

Initially, we will consider a large scale perturbation due to contributions from modes spanning only 1 e-fold - and so therefore, the variance of the large background perturbations is equal to that of the small scale perturbations, $\sigma_{l}=\sigma_{s}$. The constraints are obtained by numerically solving equation (\ref{quadratic beta}) and allowing $f_{NL}$ to vary. The results are shown in Fig.~\ref{fnl constraints} for $\beta=10^{-5}$ and $\beta=10^{-20}$. We now note that, whilst the constraints still weaken slightly for small negative values of $f_{NL}$, the constraints become tighter again as $f_{NL}$ becomes more negative, quickly becoming similar to the gaussian case - which was not seen in previous calculations \cite{Byrnes:2012yx,Young:2013oia} which neglected the long-short coupling (and hence are only valid if the power spectrum has a narrow peak). As $\lvert f_{NL}\rvert$ becomes large, the constraints asymptote to a constant value (which will be calculated in the next section).

Depending on the value of $\zeta_{l}$ in a given region of the universe, the production of PBHs can either be increased or decreased. However, the presence of large scale perturbations always increases the total number of PBHs forming in the entire universe - meaning that the power spectrum can be constrained to a lower value so that PBHs are not overproduced. This can be demonstrated by considering what happens when $f_{NL}$ is negative - it was previously found that constraints become rapidly weaker when $f_{NL}$ is negative (where the large scale background perturbations were not considered). This is due to the shape of the pdf of $\zeta$, which has a maximum value of $\zeta$ given by
\begin{equation}
\zeta_{max}=-\frac{5}{6f_{NL}}+\frac{3}{5}f_{NL}\left(\frac{25}{36f_{NL}^{2}}-\sigma^{2}\right).
\end{equation}
Unless there is fine tuning of the (local) power spectrum, this typically means that if $\sigma$ is small then no PBHs are formed, but above a critical value then so many PBHs form that the universe becomes dominated by them. However, in any given region, $\tilde{\sigma}$ and $\tilde{f}_{NL}$ are functions of $\zeta_{l}$. Therefore, depending on the value of $\zeta_{l}$, PBH production in a region can be either increased dramatically or reduced to zero. Overall, more PBHs would be produced in a universe containing such large scale inhomogeneities - and so the power spectrum is more tightly constrained. A similar but less dramatic phenomenon occurs for positive $f_{NL}$ - meaning the power spectrum can be more tightly constrained for both positive and negative $f_{NL}$.

\begin{figure}[t]
\centering
	\includegraphics[width=0.7\linewidth]{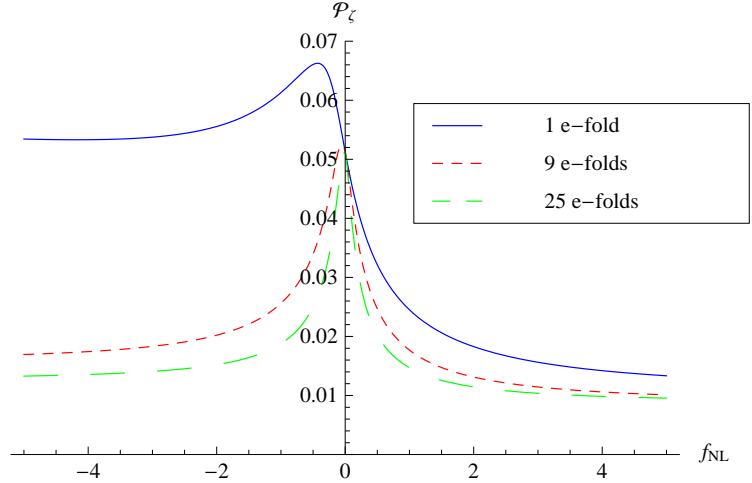}
 \caption{The constraints which can be placed on the power spectrum from PBHs depend strongly on both the amount of non-gaussianity and the amplitude of the background perturbations, given by $\langle\zeta_{l}^{2}\rangle=\mathcal{N}\mathcal{P}_{\zeta}$. This figure shows the constraint on $\mathcal{P}_{\zeta}$ from $\beta<10^{-5}$ as a function of $f_{NL}$ for $\mathcal{N}=1$, $9$ and $25$.}
 \label{many efolds quadratic}
\end{figure}

We will now consider what happens when a larger number of e-folds are considered to contribute to the background perturbation. In Fig.~\ref{many efolds quadratic} we show how the constraints change with the the variance of the background perturbations, considering the cases where the background is comprised from 9 e-folds, $\sigma_{l}=3\sigma_{s}$, and 25 e-folds, $\sigma_{l}=5\sigma_{s}$. If $f_{NL}$ is non-zero, the constraints on the small scales become much tighter as the variance on large scales increases. In order to explain this behaviour, it is useful to consider the case of large $f_{NL}$ where the linear term is dominated by the quadratic term in equation (\ref{quadratic local model}).

\subsection{Large $f_{NL}$}

If $f_{NL}$ becomes large enough such that the quadratic term dominates the linear term, we can simplify the expression for $\zeta$ to
\begin{equation}
\zeta=\pm\left(\zeta_{G}^{2}+\langle\zeta_{G}^{2}\rangle\right),
\label{quadraticNG}
\end{equation}
and performing the peak-background split as before, dropping the terms independent of $\zeta_{s}$, gives
\begin{equation}
\zeta=2\zeta_{l}\zeta_{s}\pm\left(\zeta_{s}^{2}+\sigma_{s}^{2}\right).
\end{equation}
Rewriting in terms of the variables one would observe locally
\begin{equation}
\tilde{\zeta}_{G}=2\zeta_{l}\zeta_{s},
\end{equation}
\begin{equation}
\tilde{\sigma}_{G}=2\zeta_{l}\sigma_{s},
\end{equation}
\begin{equation}
\tilde{f}_{NL}=\pm\frac{5}{12\zeta_{l}^{2}},
\end{equation}
which gives as before, see equation (\ref{localQuadInhomo}), 
\begin{equation}
\tilde{\zeta}=\tilde{\zeta}_{G}+\frac{3}{5}\tilde{f}_{NL}\left(\tilde{\zeta}_{G}^{2}-\tilde{\sigma}^{2}\right).
\end{equation}
However, we now note that, because the PDF of $\tilde{\zeta}_{G}$ is constant under a change of sign of $\tilde{\zeta}_{G}$, then the PDF of $\zeta$ is independent of the sign of $\zeta_{l}$. This can then be inserted as before into equation (\ref{quadratic beta}), which can then be solved numerically to find an upper limit on the power spectrum - this is the value that the constraints asymptote to in Fig.~\ref{fnl constraints} or \ref{many efolds quadratic}. Because the variance of the background depends on the number of e-folds it is comprised of, the constraints on the power spectrum depend on the number of e-folds between the horizon scale during PBH formation and the largest scale on which the power spectrum is enhanced, $\mathcal{N}$, see Fig.~\ref{powerSpectrum}.

\begin{figure}[t]
\centering
	\includegraphics[width=0.7\linewidth]{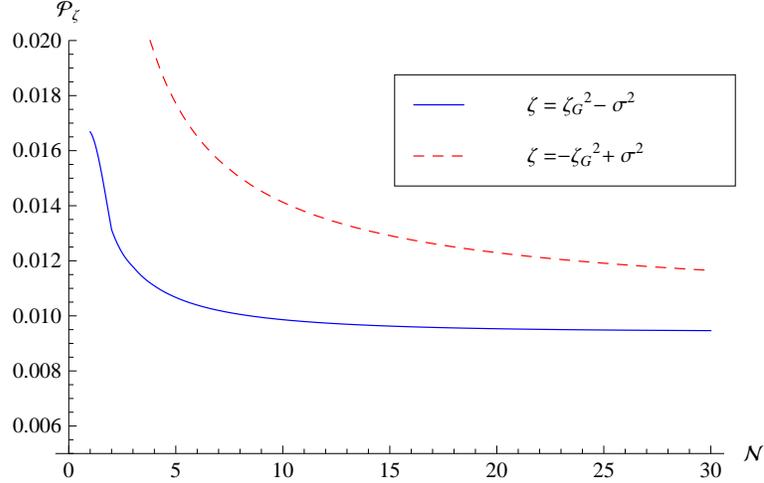}
 \caption{The constraints on the power spectrum for the quadratic case, $\zeta\approx\zeta_{G}^{2}$, are shown for $\beta<10^{-5}$ as a function of the number of e-folds of fourier modes, $\mathcal{N}$, making up the background perturbation, with $\langle\zeta_{l}^{2}\rangle=\mathcal{N}\mathcal{P}_{\zeta}$. For small $\mathcal{N}$ the constraints are much weaker for the negative case than for the positive case, and both tighten significantly as $\mathcal{N}$ becomes large. As $\mathcal{N}$ becomes very large, both will eventually asymptote to the same constant value, $\mathcal{P}_{\zeta}<9.8 \times10^{-3}$.}
 \label{quadratic limit}
\end{figure}

Fig.~\ref{quadratic limit} shows how the constraints become tighter as more e-folds are considered. For a small number of e-folds, so that $\langle\zeta_{l}^{2}\rangle$ is not too large, the constraints are much weaker for the negative quadratic case. However, as more e-folds are considered, the constraints become much closer - this is because, in universes where $\langle\zeta_{l}^{2}\rangle$ is large, then $\tilde{f}_{NL}=\pm\frac{5}{12\zeta_{l}^{2}}$ is typically small. One can therefore approximate $\tilde{\zeta}$ as gaussian\footnote{Surprisingly, starting from a completely non-gaussian distribution with a large-scale non-gaussian background, the small scales appear almost gaussian (although even small amounts of non-gaussianity have a very large effect on $\beta$). See \cite{Nelson:2012sb} for further reading.} - and the sign of the quadratic term in equation (\ref{quadraticNG}) is unimportant. Even for the weakest constraints on the abundance of PBHs, $\beta<10^{-5}$, the constraints on the power spectrum drop to $\mathcal{P}_{\zeta}<\mathcal{O}(10^{-2})$, around 5 times tighter than for the gaussian case, and 2 orders of magnitude tighter for $f_{NL}<0$ compared to when modal-coupling is not considered.

Rather than being purely hypothetical, there are models which predict such a distribution. For example $\zeta=-(g^{2}-\langle g^{2}\rangle)$ (with $g$ a gaussian variable) could be expected from the linear era of the hybrid inflation waterfall \cite{Lyth:2012yp}. The power spectrum in this model is expected to become large on some small scale before inflation ends, and peak at some value before decreasing again. In addition, $\zeta=g^{2}-\langle g^{2}\rangle$ could be predicted from a curvaton-type scenario (e.g.~\cite{Suyama:2008nt,Bugaev:2012ai,Peloso:2014oza}).

\section{Inhomogeneous cubic non-gaussianity}

The local model of non-gaussianity with a cubic term (assuming $f_{NL}=0$) is given by
\begin{equation}
\zeta=\zeta_{G}+\frac{9}{25}g_{NL}\zeta_{G}^{3}.
\label{cubicNG}
\end{equation}
We again use the peak-background split, $\zeta=\zeta_{s}+\zeta_{l}$, such that
\begin{equation}
\zeta=\left(1+\frac{27}{25}g_{NL}\zeta_{l}^{2}\right)\zeta_{s}+\left(\frac{27}{25}g_{NL}\zeta_{l}\right)\zeta_{s}^{2}+\left(\frac{9}{25}g_{NL}\right)\zeta_{s}^{3}+\mathcal{O}(\zeta_{l}),
\end{equation}
where again, the terms dependant only on $\zeta_{l}$ are unimportant in the context of PBH formation, and are neglected. $\zeta_{l}$ appears constant in a small patch of the universe, and this can be rewritten in terms of $\tilde{\zeta}_{G}$, $\tilde{\sigma}$, $\tilde{f}_{NL}$ and $\tilde{g}_{NL}$.
\begin{equation}
\tilde{\zeta}_{G}=\left(1+\frac{27}{25}g_{NL}\zeta_{l}^{2}\right)\zeta_{s},
\end{equation}
\begin{equation}
\tilde{\sigma}=\left(1+\frac{27}{25}g_{NL}\zeta_{l}^{2}\right)\sigma_{s},
\end{equation}
\begin{equation}
\tilde{f}_{NL}=\left(\frac{9}{5}g_{NL}\zeta_{l}\right)\left(1+\frac{27}{25}g_{NL}\zeta_{l}^{2}\right)^{-2},
\end{equation}
\begin{equation}
\tilde{g}_{NL}=g_{NL}\left(1+\frac{27}{25}g_{NL}\zeta_{l}^{2}\right)^{-3}.
\label{bar gnl}
\end{equation}
Therefore, equation (\ref{cubicNG}) can be rewritten as
\begin{equation}
\zeta=\tilde{\zeta}_{G}+\frac{3}{5}f_{NL}\left(\tilde{\zeta}_{G}^{2}-\tilde{\sigma}^{2}\right)+\frac{9}{25}\tilde{g}_{NL}\tilde{\zeta}_{G}^{3},
\end{equation}
where the $-\tilde{\sigma}^{2}$ term has been inserted manually to ensure $\langle\zeta\rangle=0$. An expression for the abundance of PBHs in a given region of the universe, $\tilde{\beta}$, can be derived in terms of $\tilde{\sigma}$, $\tilde{f}_{NL}$ and $\tilde{g}_{NL}$ - see \cite{Byrnes:2012yx} for details, we do not give the full calculation here. Again, in order to derive the complete expression for the abundance of PBHs in the entire universe, it is necessary to integrate over $\zeta_{l}$ as before,
\begin{equation}
\beta=\int_{-\infty}^{\infty}\tilde{\beta}(\zeta_{l})P(\zeta_{l})d\zeta_{l}.
\end{equation}
This expression can then be solved numerically to derive a constraint on $\sigma$ from a constrain on $\beta$. The constraint on the power spectrum, $\mathcal{P}_{\zeta}$ can then be calculated using \cite{Byrnes:2007tm}
\begin{equation}
\mathcal{P}_{\zeta}=\sigma^{2}+6\left(\frac{9g_{NL}}{25}\right)\sigma^{4}\ln(kL)+27\left(\frac{9g_{NL}}{25}\right)^{2}\sigma^{6}\ln(kL)^{2}.
\end{equation}

\begin{figure}[t]
\centering
\begin{subfigure}{0.5\textwidth}
	\centering
	\includegraphics[width=\linewidth]{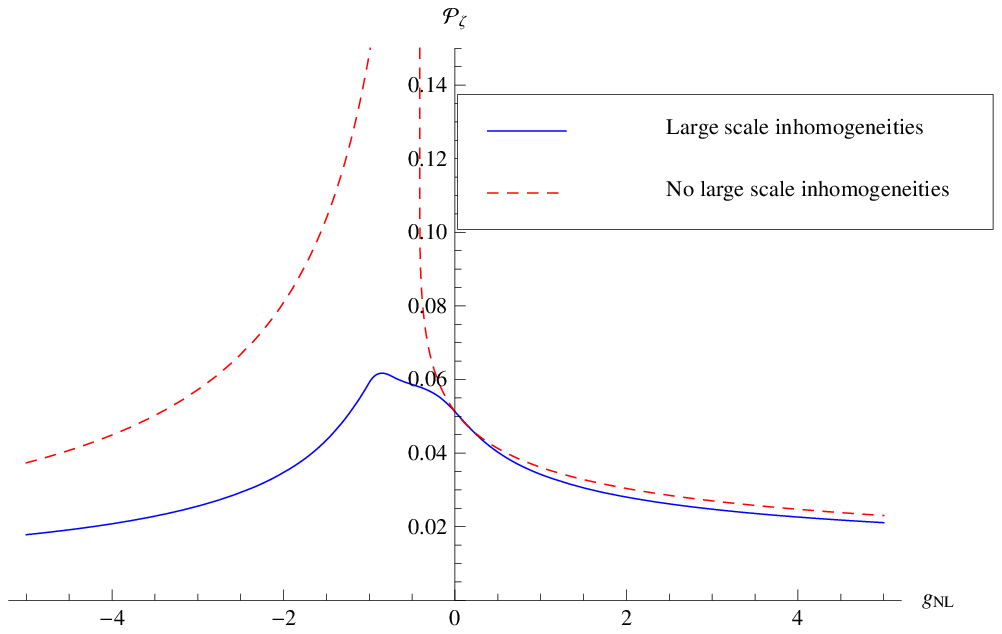}
	\caption{$\beta=10^{-5}$}
\end{subfigure}%
\begin{subfigure}{0.5\textwidth}
	\centering
	\includegraphics[width=\linewidth]{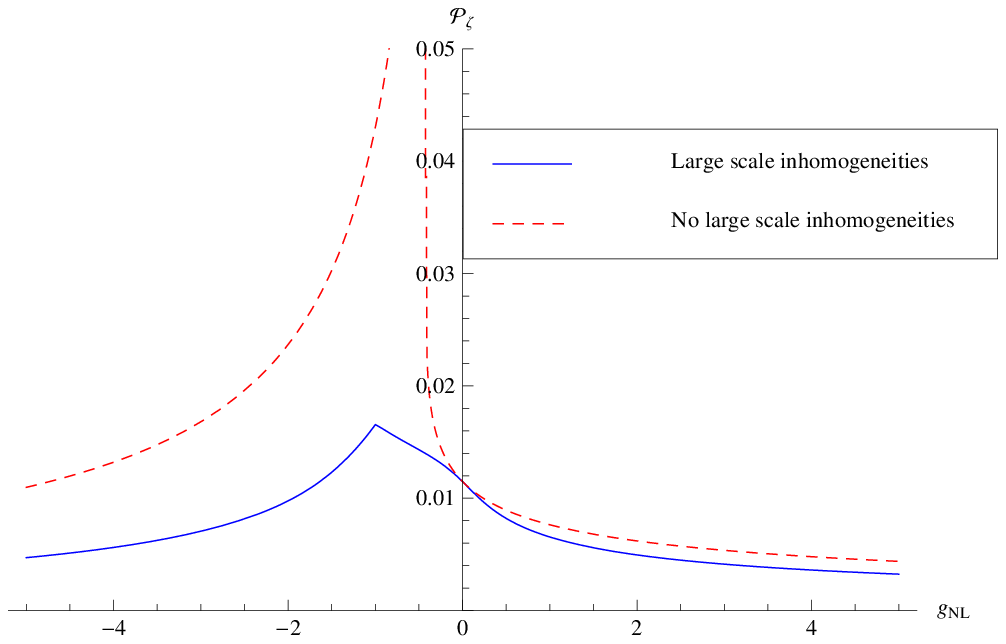}
	\caption{$\beta=10^{-20}$}
\end{subfigure}
\caption{The constraints that can be placed upon the power spectrum are displayed - they depend significantly upon the value of the non-gaussianity parameter, $g_{NL}$. The dotted red lines show the constraints calculated previously, where the effect of large scale modes was not considered, and the solid blue lines show the constraints when they are included. The left plot (a) displays the constraints for $\beta<10^{-5}$ and the right plot (b) for $\beta<10^{-20}$. In this plot it is assumed that the variance of the gaussian component of the large scale perturbations is the same as that of the small scale perturbations, $\langle\zeta_{l}^{2}\rangle=\langle\zeta_{s}^{2}\rangle=\sigma^{2}$. Typically, the constraints tighten significantly when there is any non-gaussianity present - with a slight weakening for small negative $g_{NL}$. The constraints are significantly tighter than previously calculated, and do not display as sharp a peak for small negative $g_{NL}$ where the constraints became rapidly weaker.}
\label{gnlCons}
\end{figure}

Fig.~\ref{gnlCons} shows how the constraints on the power spectrum depend on $g_{NL}$ for $\beta=10^{-5}$ and $\beta=10^{-20}$. Again, we see that constraints become tighter as the non-gaussianity parameter $g_{NL}$ becomes large. However, the sharp peak seen in previous calculations is now smoothed out, and the constraints are significantly tighter - this is because only for a small range of values of $g_{NL}$ is the production of PBHs significantly reduced (seen by the region in which the constraints weaken in Fig.~\ref{old constraints}), but the background perturbations cause $g_{NL}$ to vary, see equation (\ref{bar gnl}). As seen in previous papers, as $\lvert g_{NL}\rvert$ becomes large, the constraints asymptote to the same value for negative or positive $g_{NL}$ - which is as expected (this will be explored in the next section).

\begin{figure}[t]
\centering
	\includegraphics[width=0.7\linewidth]{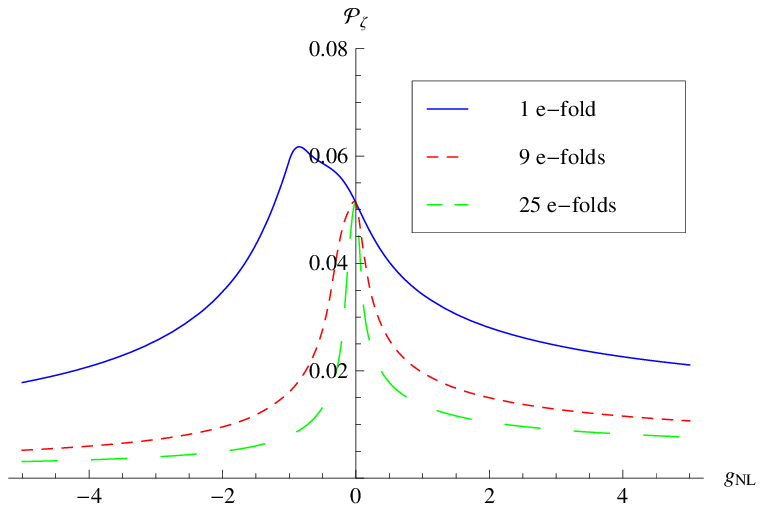}
\caption{As in the quadratic case, the constraints which can be placed on the power spectrum from PBHs depend strongly on both the amount of non-gaussianity and the amplitude of the background perturbations, given by $\langle\zeta_{l}^{2}=\mathcal{NP}_{\zeta}$. This figure shows the constraint on $\mathcal{P}_{\zeta}$ from $\beta < 10^{-5}$ as a function of $g_{NL}$ for $\mathcal{N} =1$, $9$ and $25$, becoming much tighter as more e-folds are considered.}
\label{efolds gnl}
\end{figure}

We will now again consider the constraints if the background perturbations consist of multiple e-folds of perturbations. Fig.~\ref{efolds gnl} shows the resultant constraints obtained if the background perturbations consist of 1, 9, or 25 e-folds, as before. When more e-folds are considered, the constraints become much tighter - only for small negative $g_{NL}$ do the constraints weaken slightly, but for all other values of $g_{NL}$ the constraints become significantly tighter, $\mathcal{P}_{\zeta}<\mathcal{O}(10^{-3})$ for even small values of $g_{NL}$.

\subsection{Large $g_{NL}$}

\begin{figure}[t]
\centering
	\includegraphics[width=0.7\linewidth]{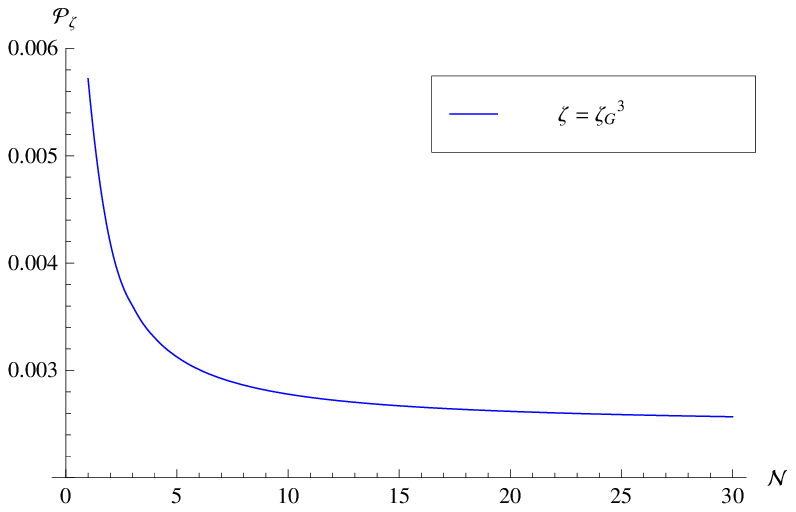}
 \caption{The constraints on the power spectrum for the cubic case, $\zeta\approx\pm\zeta_{G}^{3}$, are shown for $\beta<10^{-5}$ as a function of the number of e-folds of fourier modes, $\mathcal{N}$, making up the background perturbation, with $\langle\zeta_{l}^{2}\rangle=\mathcal{NP}_{\zeta}$. Similar to the quadratic case, the constraints tighten significantly as the number of e-folds being considered increases, eventually reaching a constant for large $\mathcal{N}$ at $\mathcal{P}_{\zeta}<2.5\times10^{-3}$.}
\label{cubic limit}
\end{figure}

We will now consider the case where the cubic term dominates, and $\zeta$ can be expressed as
\begin{equation}
\zeta_{\pm}=\pm \zeta_{G}^{3}.
\end{equation}
In the cubic case, the sign does not matter - because a gaussian distribution is symmetric, the PDF of $\zeta_{+}$ and $\zeta_{-}$ is the same, and we will therefore drop the dependance on the sign and discuss only the positive case. Completing the peak-background split and isolating the short scale gives
\begin{equation}
\zeta=3\zeta_{l}^{2}\zeta_{s}+3 \zeta_{l}\left(\zeta_{s}^{2}-\sigma_{s}^{2}\right)+\zeta_{s}^{3},
\end{equation}
where we have inserted the $\sigma_{s}^{2}$ term manually. Again, defining effective short-scale parameters:
\begin{equation}
\tilde{\zeta}_{G}=3\zeta_{l}^{2}\sigma_{s},
\end{equation}
\begin{equation}
\tilde{\sigma}=3\zeta_{l}^{2}\sigma_{s},
\end{equation}
\begin{equation}
\tilde{f}_{NL}=\frac{5}{3}\left(3\zeta_{l}^{2}\right)^{-2},
\end{equation}
\begin{equation}
\tilde{g}_{NL}=\frac{25}{9}\left(3\zeta_{l}^{2}\right)^{-3}.
\end{equation}
We note that as $\zeta_{l}$ becomes large, the small scale observable universe will appear more gaussian. The constraints on the power spectrum $\mathcal{P}_{\zeta}$ can then be computed numerically as before from constraints on the mass fraction of PBHs $\beta$, as a function of the number of e-folds considered in the background perturbation, $\mathcal{N}$ - the results can be seen in Fig.~\ref{cubic limit}. We see that, for a moderate number of e-folds considered, the constraints drop to $\mathcal{P}_{\zeta}<\mathcal{O}(10^{-3})$, eventually tightening to $\mathcal{P}_{\zeta}<2.4\times10^{-3}$.

\section{Conclusions}
We have extended the calculation for the abundance of PBHs, defined in terms of the mass fraction of the universe forming PBHs at the time of formation $\beta$, when there is non-gaussianity present to include the effect of coupling between large scale super-horizon modes and smaller horizon scale perturbations. We see that non-gaussianity typically increase the overall amount of PBHs which would form - with some regions of the universe producing significantly more PBHs than other regions. A realisation of such a universe - containing significant non-gaussianity and a broad peak in the power spectrum at scales significantly smaller than those visible in the CMB is possible in hybrid inflation, and in particular from the waterfall transition of $\mathcal{N}$-field hybrid inflation \cite{Halpern:2014mca}.

Observational constraints on $\beta$, which range from $\beta<10^{-5}$ to $\beta<10^{-20}$, can then be used to place an upper constraint on the primordial curvature perturbation power spectrum, $\mathcal{P}_{\zeta}$. We have investigated the constraints which can be placed on the power spectrum dependant on the amount of non-gaussianity present and the coupling between modes, for a simple model of the power spectrum. Because non-gaussianity typically increases PBH formation, the constraints on $\mathcal{P}_{\zeta}$ are typically much tighter - and we show that the constraints from PBHs may be significantly tighter than calculated in previous work. The presence of non-gaussianity and large super-horizon modes have a large impact on the constraints - and when there is significant non-gaussianity the constraints can become tighter by several orders of magnitude. The effect of simultaneously having a non-zero $f_{NL}$ and $g_{NL}$ have also been considered, although the analysis has not been explicitly included in this paper. It is again found that small negative values of $f_{NL}$ or $g_{NL}$ weaken the constraints slightly, but typically the constraints become stronger.

In this paper, we have considered local-type (squeezed) non-gaussianity, which includes a significant coupling between the modes \cite{Komatsu:2009kd}. We would expect results to be similar for flattened-type non-gaussianity as there is still a significant coupling between modes of different lengths (albeit weaker than in the local model). However, for equilateral type non-gaussianity (which is peaked in the limit of all three modes having the same wavelength) we would not expect significant coupling between large and short scales, so the results would be expected to more closely reflect previous analyses in which large amplitude perturbations on only one scale were considered. However there have not been any detailed studies made of how non-Gaussianity of non local shapes effects the bounds on PBHs.

The main source of error in the calculation arises from the uncertainty in the formation criterion, which lies in the range $0.7<\zeta_{c}<1.2$ - and this has a very large effect on the calculated value for $\beta$, which can easily vary by several orders of magnitude \cite{Young:2014ana}. However, the effects on the constraints calculated are much less drastic, and the error due to the uncertainty in $\zeta_{c}$ is expected to be of order $10\%$. There is also uncertainty of how intermediate modes should be handled, which are currently excluded from the calculation - how long does a mode have to be before it is considered to be part of the background? The size of this cut-off scale can have a non-negligible effect on the constraints calculated - although how important the effect is depends on the specific form of the power spectrum being considered. In this paper, we have avoided this uncertainty by considering the background perturbations to result from a given number of e-folds of modes.

We also note that the Taylor-type expansion of $\zeta$ in terms of $f_{NL}$ and $g_{NL}$, which we have used here, may not give an accurate result for the constraints. It was shown in a previous paper \cite{Young:2013oia} that higher orders terms can have a significant effect, and care should therefore be taken to ensure that results are valid when calculating constraints for a specific model.

\section{Acknowledgements}
The authors would like to thank Misao Sasaki for useful discussion during the production of this paper. 
SY is supported by an STFC studentship, and would like to thank Yukawa Institute for Theoretical Physics for its hospitality during a month long stay which was supported by the Bilateral International Exchange Program (BIEP). CB is supported by a Royal Society University Research Fellowship.

\bibliographystyle{JHEP}
\bibliography{bibfile}

\end{document}